\def\kF{k_{\text{F}}}
\def\kB{k_{\text{B}}}
\def\NF{N_{\text{F}}\,}
\def\me{m_{\text{e}}}
\def\be{\begin{equation}}
\def\ee{\end{equation}}
\def\bea{\begin{eqnarray}}
\def\eea{\end{eqnarray}}
\def\bse{\begin{subequations}}
\def\ese{\end{subequations}}

\documentclass[prb,twocolumn,showpacs,preprintnumbers,amsmath,amssymb,eqsecnum]{revtex4}
\usepackage{graphicx}
\usepackage{dcolumn}
\usepackage{bm}
\begin{document}
\title{Smectons: Soft modes in electronic stripe phases, and their consequences
       for thermodynamics and transport
}
\author{T.R. Kirkpatrick$^{1}$ and D. Belitz$^{2}$}
\affiliation{$^{1}$Institute for Physical Science and Technology and Department
                   of Physics, University of Maryland, College Park, MD 20742\\
             $^{2}$Department of Physics, Institute of Theoretical Science, and
                   Materials Science Institute, University of Oregon, Eugene, OR
                   97403}
\date{\today}
\begin{abstract}
The Goldstone mode due to stripe or unidirectional charge-density-wave order in
electron systems is found to have the same functional form as the one in
classical smectic liquid crystals. It is very similar to the Goldstone mode
that results from helical magnetic order. This allows for an effective theory
that provides a quasiparticle description of either stripe phases or
helimagnets in the low-energy regime. The most remarkable observable
consequence is an electronic relaxation rate in $d=2$ that is $1/\tau\propto
T\ln T$ in clean systems and $1/\tau\propto \sqrt{T}$ in weakly disordered
ones. The corresponding results in $d=3$ are $1/\tau\propto T^{3/2}$ and
$1/\tau\propto T$, respectively.
\end{abstract}

\pacs{}

\maketitle

\section{Introduction}
\label{sec:I}

Classical liquid crystals display a fascinating variety of ordered phases.
\cite{DeGennes_Prost_1993, Chaikin_Lubensky_1995} A fluid of prolate molecules
(``directors'') can enter a nematic phase that breaks rotational invariance by
aligning, on average, the major axes of the molecules while their
center-of-mass-motion remains fluid-like. If, in addition, the molecules
arrange in layers that break translational invariance in one direction, a
smectic phase results. Smectic order is characterized by a vector ${\bm q}$
whose direction is normal to the layers, and whose modulus $q$ determines the
inter-layer distance $2\pi/q$. One further distinguishes between smectic-A
phases, where, in the ground state, the major axes of the directors are aligned
perpendicular to the layers, and smectic-C phases, where they are aligned at an
angle. If chiral molecules are added to the fluid, a cholesteric phase can
result where translational invariance is broken by means of a helical
arrangements of the directors. These various instances of spontaneously broken
continuous symmetries result in the existence of Goldstone modes.
\cite{Lubensky_1972, Martin_Parodi_Pershan_1972} In the nematic phase the
Goldstone modes correspond to uniform rotations of all molecules, analogous to
the Goldstone modes in a ferromagnet. In the smectic and cholesteric phases the
Goldstone mode takes the form of a propagating wave with a highly anisotropic
frequency-wave vector relation:
\be
\Omega_{\text{s}}({\bm k}) = \frac{\vert{{\bm k}_\perp}\vert}{\vert{\bm
k}\vert}\, \sqrt{c_x\,k_x^2 + c_\perp\,{\bm k}_\perp^4/q^2}\ .
\label{eq:1.1}
\ee
Here ${\bm k} = (k_x,{\bm k}_{\perp})$ is the wave vector, and we have chosen
${\bm q}$ to point in the $x$-direction. $c_x$ and $c_\perp$ are elastic
constants. The excitation with frequency $\Omega_{\text{s}}$ is often referred
to as ``second sound''. Ordinary, or first, sound also exists and is slightly
modified by the existence of the smectic or cholesteric order. The absence of a
term proportional to ${\bm k}_\perp^2$ under the square root in Eq.\
(\ref{eq:1.1}) is due to rotational invariance, and the functional form of
$\Omega_{\text{s}}({\bm k})$ is the same in both smectic and cholesteric
phases.

In recent years, electronic analogs of liquid-crystal ordered phases have been
discussed in the context of quantum Hall
systems,\cite{Fogler_Koulakov_Shklovskii_1996, Moessner_Chalker_1996,
Lilly_et_al_1999} high-T$_\text{c}$ superconductors,
\cite{Kivelson_Fradkin_Emery_1998} and helical
magnets.\cite{Binz_Vishvanath_Aji_2006, Tewari_Belitz_Kirkpatrick_2006,
Fischer_Shah_Rosch_2008} An electronic nematic phase can result from an
attractive electron-electron interaction in the quadrupole or $\ell=2$ channel.
\cite{Oganesyan_Kivelson_Fradkin_2001} If the interaction amplitude exceeds a
critical strength, a Pomeranchuk instability \cite{Pomeranchuk_1958} results in
a Fermi surface that, for nearly-free electrons, is an ellipse (in $d=2$) or an
ellipsoid (in $d=3$) instead of a circle or a sphere. With increasing
correlation strength, a unidirectional charge-density wave can form that is the
electronic analog of a smectic phase. Such one-dimensional order is normally
unstable, but it is stabilized by the underlying nematic order.
\cite{Fradkin_Kivelson_1999} The resulting ``stripe phases'' are believed to be
realized in quantum Hall systems and in high-T$_{\text{c}}$ superconductors.
\cite{Kivelson_et_al_2003} An electronic analog of cholesteric order is
provided by helical magnets, such as MnSi or FeGe, where the magnetization
orders in a helical pattern.\cite{Ishikawa_et_al_1976} The Goldstone mode in
the latter (``helimagnon'') turns out to be very similar to that in either
classical cholesterics or smectics; it is given by Eq.\ (\ref{eq:1.1}) without
the $\vert{\bm k}_\perp\vert/\vert{\bm k}\vert$ prefactor.
\cite{Belitz_Kirkpatrick_Rosch_2006a} This difference is due to differences in
the kinetic equations that govern the dynamics of spins and directors,
respectively.

In this paper we investigate stripe order, with a focus on $2$-$d$ or
quasi-$2$-$d$ systems, although the corresponding $3$-$d$ results can be
readily obtained and are also given. We determine the resulting Goldstone mode
and its consequences for observables. We will focus on systems in a vanishing
magnetic field; for discussions of soft fluctuations in stripe phases of
quantum Hall systems, see, Refs.\ \onlinecite{Fogler_Vinokur_2000,
Lopatnikova_et_al_2001, Wexler_Dorsey_2001}. We find that the contribution of
the Goldstone mode to the specific heat is proportional to $T^{3/2}$, and thus
subleading to the Fermi-liquid contribution. The single-particle relaxation
rate $1/\tau$ averaged over the Fermi surface, however, is found to go as $T\ln
T$, which is a much stronger $T$-dependence than in a Fermi liquid. The former
result is consistent with the one obtained before in Ref.\
\onlinecite{Sun_et_al_2008}, but the latter is not. We will explain the origin
of this discrepancy. In addition, our results are the $d=2$ analogs of our
previous results for helimagnets.\cite{Belitz_Kirkpatrick_Rosch_2006a,
Belitz_Kirkpatrick_Rosch_2006b} Our result for $1/\tau$ implies an inverse
thermal conductivity that goes as $T\ln T$, and we will discuss consequences
for the electrical conductivity.

\section{Stripe order}
\label{sec:II}

\subsection{Statics}
\label{subsec:II.A}

Let us assume a phase with stripe order, i.e., an electron density $\rho$ in
the ground state that can be written, in a saddle-point
approximation,\cite{harmonics_footnote} as
\bse
\label{eqs:2.1}
\be
\rho_{\text{\,sp}}(x) = \rho_0 + \Delta\,\cos({\bm q}\cdot{\bm x}).
\label{eq:2.1a}
\ee
Here $\rho_0$ is the average density, and $\Delta$ is the amplitude of the
density wave, which is the order parameter of the smectic order. $x=({\bm
x},\tau)$ comprises the spatial position ${\bm x}$ and the imaginary time
$\tau$. The density wave vector ${\bm q}$ with modulus $q \equiv \vert{\bm
q}\vert$ is determined by the microscopic mechanism that causes the smectic
order, e.g., a density-density correlation function that has a maximum at
$q\neq 0$. In general, in an electronic smectic one expects $q$ to be a sizable
fraction of the Fermi wave number $\kF$.

Fluctuations about the saddle point will include amplitude fluctuations, which
are massive and can be neglected, and phase fluctuations that will be soft. The
fluctuating density will thus read
\be
\rho(x) = \rho_0 + \Delta\,\cos({\bm q}\cdot{\bm x} + u(x)),
\label{eq:2.1b}
\ee
\ese
with a phase $u(x)$. The functional form of the static phase-phase correlation
function is determined by rotational symmetry and must be the same as in the
classical case,\cite{Chaikin_Lubensky_1995}
\be
\langle u(k)\,u(-k)\rangle_{i\Omega=0} = \frac{1}{\NF}\, \frac{1}{c_x\,k_x^2 +
c_{\perp}\,{\bm k}_\perp^4/q^2}\ .
\label{eq:2.2}
\ee
Here $k = ({\bm k},i\Omega)$ comprises the wave vector ${\bm k}$ and a bosonic
imaginary frequency $i\Omega$. Since any Gaussian action must be quadratic in
$\Delta$, the elastic constants will be proportional to $\Delta^2$ and
proportional to one another: $c_x \propto c_\perp \propto \lambda^2/\kF^{\!\!
2}$. Here $\lambda = \Gamma\Delta$, with $\Gamma$ an appropriate
density-density interaction strength, is an energy that is the charge-density
analog of the Stoner gap in ferromagnetism, and the density of states on the
Fermi surface, $\NF$, serves as a normalization factor.

\subsection{Dynamics}
\label{subsec:II.B}

We now use time-dependent Ginzburg-Landau theory (TDGL) \cite{Ma_1976} to
determine the dynamics of the phase fluctuations. For density fluctuations, the
appropriate kinetic equation is a Langevin equation\cite{damping_footnote}
\bse
\label{eqs:2.3}
\be
\rho\,\partial_{\,t} {\bm v} = -{\bm\nabla}p - \rho({\bm
v}\cdot{\bm\nabla})\,{\bm v} - e\,(\rho/\me)\,{\bm E} + {\bm\zeta},
\label{eq:2.3a}
\ee
augmented by the continuity equation
\be
\partial_{\,t}\,\rho = -{\bm\nabla}\cdot(\rho\,{\bm v}),
\label{eq:2.3b}
\ee
and the Maxwell equations\cite{Poisson_footnote}
\bea
{\bm\nabla}\cdot{\bm E} &=& -2^{d-1}\pi\,e\,(-{\bm\nabla}^2)^{(3-d)/2}\,(\rho -
\rho_0).
\label{eq:2.3c}\\
{\bm\nabla}\times{\bm E} &=& 0.
\label{eq:2.3d}
\eea
\ese
Here $t$ denotes real time, $\rho$ is the mass density, ${\bm v}$ is the
velocity, $\me$ and $e$ are the electron mass and charge, respectively, ${\bm
E}$ is the electric field, and ${\bm\zeta}$ is a Langevin force. The pressure
$p$ can be written $p = (\rho/V)\partial F/\partial\rho$, with $V$ the system
volume and $F$ the free energy. The latter is given by a Hamiltonian $H[u]$
that generates the static correlation function given by Eq.\ (\ref{eq:2.2}). At
this point we need to realize that, because of the conserved nature of the
density, fluctuations near ${\bm k}=0$ are as important as those near ${\bm k}
= {\bm q}$. We thus write, to linear order in the fluctuations,
\bse
\label{eqs:2.4}
\be
\rho({\bm x}) = \rho_0 + n({\bm x}) + \Delta\,\cos({\bm q}\cdot{\bm x}) -
u({\bm x})\,\Delta\,\sin({\bm q}\cdot{\bm x}),
\label{eq:2.4a}
\ee
and
\bea
H[n,u] &=& (c_0^2/2\rho_0)\int d{\bm x}\ \left(n({\bm x})\right)^2
\nonumber\\
&&\hskip -30pt + \NF \int d{\bm x}\ \left[c_x\left(\partial_x\,u({\bm
x})\right)^2 + c_\perp\left({\bm \nabla}_{\perp}^2 u({\bm x})\right)^2\right],
\nonumber\\
\label{eq:2.4b}
\eea
\ese
with $c_0$ the speed of (first) sound. The Eqs.\ (\ref{eqs:2.3}) are now fully
specified, and we solve them in a zero-loop approximation, which amounts to
neglecting the terms nonlinear in the velocity.\cite{Hohenberg_Halperin_1977}
Within TDGL, the static fields entering the Hamiltonian are replaced by the
corresponding dynamic ones after performing the appropriate functional
derivatives.

We now course grain the equations, i.e., we perform a spatial average over a
small volume that contains an integer number of charge density wave periods.
This makes the Langevin force in Eq.\ (\ref{eq:2.3a}) drop out, and we obtain
\bse
\label{eqs:2.5}
\bea
\partial_{\,t} {\bm v}({\bm x},t) &=& \frac{-c_0^2}{\rho_0}\, {\bm\nabla} n({\bm
x},t) + \frac{{\bm q}}{q^2}\,\omega_0^2({\bm\nabla})\, u({\bm x},t)
\nonumber\\
&&\hskip 50pt - \frac{e}{\me}\,{\bm E}({\bm x},t).
\label{eq:2.5a}
\eea
Here we have used the identities
\[
\frac{\delta H}{\delta\rho({\bm x})} = \int d{\bm y}\ \left(\frac{\delta
H}{\delta u({\bm y})}\, \frac{\delta u({\bm y})}{\delta\rho({\bm x})} +
\frac{\delta H}{\delta n({\bm y})}\, \frac{\delta n({\bm y})}{\delta\rho({\bm
x})}\right),
\]
with $H$ from Eq.\ (\ref{eq:2.4b}), and
\[
\delta n({\bm y})/\delta\rho({\bm x}) = \delta({\bm x}-{\bm y}),
\]
and defined an operator
\be
\omega_0^2({\bm\nabla}) = \gamma_0\,[-c_x\,\partial_x^2 +
c_{\perp}\,{\bm\nabla}_\perp^4/q^2]
\label{eq:2.5b}
\ee
with $\gamma_0 = 2\NF q^2/\rho_0$. For later reference we also define
\be
\omega_{\text{p}}^2({\bm\nabla}) = 2^{d-1} \pi\,\frac{\rho_0 e^2}{\me^2}\,
(-{\bm\nabla}^2)^{(3-d)/2} - c_0^2 {\bm\nabla}^2.
\label{eq:2.5c}
\ee
\ese
Course graining the continuity equation, and using Eq.\ (\ref{eq:2.4a}),
yields
\bse
\label{eqs:2.6}
\be
\partial_{\,t} n({\bm x},t) = -\rho_0\,{\bm\nabla}\cdot{\bm v}({\bm x},t).
\label{eq:2.6a}
\ee
Also from the continuity equation, by multiplying by $\sin ({\bm q}\cdot{\bm
x})$ and coarse graining, we find
\be
\partial_{\,t} u({\bm x},t) = -{\bm q}\cdot{\bm v}({\bm x},t).
\label{eq:2.6b}
\ee
\ese
Finally, from Eqs.\ (\ref{eq:2.3c}, \ref{eq:2.3d}) we obtain, after coarse
graining,
\be
{\bm\nabla}^2 {\bm E}({\bm x},t) =
-2^{d-1}\pi\,e\,(-{\bm\nabla}^2)^{(3-d)/2}\,n({\bm x},t).
\label{eq:2.7}
\ee

We next derive an equation that couples $n$ and $u$ by taking the divergence of
Eq.\ (\ref{eq:2.5a}), and using Eqs.\ (\ref{eq:2.3c}) and (\ref{eq:2.6a}):
\bse
\label{eqs:2.8}
\bea
\partial_{\,t}^2 n({\bm x},t) &=& -\omega_{\text{p}}^2({\bm\nabla})\, n({\bm x},t)
\nonumber\\
&&- \frac{\rho_0}{q^2}\,({\bm q}\cdot{\bm\nabla})\,\omega_0^2({\bm\nabla})\,
u({\bm x},t).
\label{eq:2.8a}
\eea
A second equation coupling $n$ and $u$ is obtained from Eq.\ (\ref{eq:2.6b})
with the help of Eqs.\ (\ref{eq:2.5a}) and (\ref{eq:2.7}), viz.
\bea
\partial_{\,t}^2 {\bm\nabla}^2 u({\bm x},t) &=& -{\bm\nabla}^2
\omega_0^2({\bm\nabla})\, u({\bm x},t)
\nonumber\\
&& - \frac{q}{\rho_0}\ \omega_{\text p}^2({\bm\nabla})\,
\partial_{\,t} n({\bm x},t).
\nonumber\\
\label{eq:2.8b}
\eea
\ese

The equations (\ref{eqs:2.8}) constitute a closed system of partial
differential equations for the two dynamical variables $n$ and $u$. With
\bse
\label{eqs:2.9}
\be
\omega_0^2({\bm k}) = \gamma_0\,[c_x\,k_x^2 + c_{\perp}\,{\bm k}_\perp^4/q^2]
\label{eq:2.9a}
\ee
the Fourier transform of Eq.\ (\ref{eq:2.5b}), and
\be
\omega_{\text{p}}^2({\bm k}) = 2^{d-1}\pi\,\frac{\rho_0\,e^2}{\me^2}\,
\vert{\bm k}\vert^{3-d} + c_0^2\,{\bm k}^2
\label{eq:2.9b}
\ee
\ese
the plasma frequency squared, the two resonance frequencies are
\bse
\label{eqs:2.10}
\bea
\Omega_{\text{p}}^2({\bm k}) &=& \omega_{\text{p}}^2({\bm k}) + \omega_0^2({\bm
k})\,k_x^2/{\bm k}^2,
\label{eq:2.10a}\\
\Omega_{\text{s}}^2({\bm k}) &=& \omega_0^2({\bm k})\,{\bm k}_{\perp}^2/{\bm
k}^2.
\label{eq:2.10b}
\eea
\ese
For neutral systems ($e=0$), $\omega_{\text{p}}({\bm k})$ correctly reduces to
the frequency of (first) sound, $c_0\,\vert{\bm k}\vert$. Switching back to an
imaginary time/frequency representation, the corresponding eigenvectors, i.e.,
the soft modes, are
\bse
\label{eqs:2.11}
\bea
\pi(k) &=& n(k) + i\,\frac{\rho_0}{qc_0^2}\, \omega_0^2({\bm k})\,
\frac{k_x}{{\bm k}^2}\, u(k),
\label{eq:2.11a}\\
\sigma(k) &=& u(k) + i\, \frac{q}{\rho_0}\, \frac{k_x}{{\bm k}^2}\, n(k).
\label{eq:2.11b}
\eea
\ese
The $\sigma$-$\sigma$ correlation function or smecton susceptibility then is
\be
\chi_{\sigma\sigma}(k) = \frac{\gamma_0\, q^2}{\NF\kF^{\!\! 2}}\
\frac{1}{\Omega_{\text{s}}^2({\bm k}) - (i\Omega)^2}\ .
\label{eq:2.12}
\ee
These results are exactly analogous to the classical case,\cite{Lubensky_1972,
Martin_Parodi_Pershan_1972} see Eq.\ (\ref{eq:1.1}), except that the charged
nature of the electron system modifies first sound into a plasma mode. Note the
strongly anisotropic wave-vector dependence of both resonance frequencies, and
of $\Omega_{\text{s}}$ in particular. In a classical context, the two modes are
usually referred to as first and second sound, respectively, for neutral
systems, and as plasma oscillations and second sound for charged ones. In a
quantum context, the quanta of first sound and plasma oscillations are usually
referred to as phonons and plasmons, respectively, and by analogy we call the
quanta of second sound `smectons'. The smecton is the Goldstone mode related to
the unidirectional charge-density wave order. It is precisely analogous to the
corresponding soft modes in both smectic and cholesteric liquid crystals. By
contrast, the helimagnon (the Goldstone mode in a helical magnet), is {\em not}
completely analogous to the classical cholesteric Goldstone mode; it is missing
the factor ${\bm k}_{\perp}^2/{\bm k}^2$ in the analog of Eq.\
(\ref{eq:2.10b}).\cite{Belitz_Kirkpatrick_Rosch_2006a} This factor is also
missing in the phenomenological description of an electronic smectic in Ref.\
\onlinecite{Sun_et_al_2008}, which did not take into account the coupling to
plasmons.\cite{soft_mode_footnote} While this omission makes a large difference
for the angular dependence of the resonance frequency, it is of no consequence
for the leading temperature dependencies of various observables that couple to
the smectons. This is because, as we will see, the latter is determined by a
region in wave-vector space where $k_x \sim {\bm k}_{\perp}^2 \sim T$ in a
scaling sense,\cite{notation_footnote} and in this regime the prefactor is
equal to unity to leading order as $T\to 0$.

\section{Observable consequences of the smectons}
\label{sec:III}

\subsection{Specific heat}
\label{subsec:III.A}

A result that can be immediately obtained from the resonance frequency alone is
the smecton contribution to the internal energy $U$, and hence to the specific
heat $C = \partial U/V\partial T$. The former is given by $U_{\text{s}} =
\sum_{\bm k} \Omega_{\text{s}}({\bm k})\, n_{\text{B}}(\Omega_{\text{s}}({\bm
k}))$, with $n_{\text B}$ the Bose distribution function. The result in $3$-$d$
is $C(T\to 0) \propto T^2$, and in $2$-$d$ we find
\be
C_{\text{s}}(T\to 0) = A_C\,q^2\,(T/T_q)^{3/2}.
\label{eq:3.1}
\ee
Here we have defined a temperature scale $T_q =
\sqrt{\gamma_0}\lambda\,q^2/\kF^2$, and $A_C$ is a number of $O(1)$. This
agrees with the result obtained in Ref.\ \onlinecite{Sun_et_al_2008}. An
inspection of the integral shows that the dominant contribution comes from the
region in wave-vector space mentioned at the end of the last section. The
resulting temperature dependence is non-analytic, but subleading compared to
the Fermi-liquid contribution $C_{\text{FL}} \propto T$.

\subsection{Effective quasiparticle theory}
\label{subsec:III.B}

In order to consider the effects of the smectons on other observables it is
useful to derive an effective action for quasiparticles in the presence of
smectic order, in analogy to the theory for helical magnets developed in Ref.\
\onlinecite{Kirkpatrick_Belitz_Saha_2008a}. In the present case, the effective
action takes the form
\bse
\label{eqs:3.2}
\be
S_{\text{eff}}[\bar\psi,\psi] = S_0[\bar\psi,\psi] + \frac{1}{2}\,\Gamma^2 \int
dx\,dy\ n_q(x)\,\chi(x,y)\,n_q(y).
\label{eq:3.2a}
\ee
Here $\bar\psi$ and $\psi$ are fermion fields, $\Gamma$ is the interaction
amplitude mentioned in the context of Eq.\ (\ref{eq:2.2}) above, and $n_q(x)$
is the electron number density $\bar\psi(x) \psi(x)$ with the understanding
that $n_q$ contains only wave vectors close to ${\bm q}$ or $-{\bm q}$. $S_0$
is given by
\be
S_0[\bar\psi,\psi] = {\tilde S}_0[\bar\psi,\psi] + \lambda \int dx\ \cos({\bm
q}\cdot{\bm x})\, n_q(x),
\label{eq:3.2b}
\ee
and ${\tilde S}_0$ describes free or band electrons plus any interactions in
channels other than the one mediated by $\Gamma$. $\chi$ is the density
susceptibility in the relevant wave-vector region, which is dominated by the
phase-phase susceptibility $\chi_{uu}$, which in turn, as far as leading
hydrodynamic effects are concerned, is the same as the smecton susceptibility,
Eq.\ (\ref{eq:2.12}):
\be
\chi(x,y) \approx \Delta^2\,\sin({\bm q}\cdot{\bm x})\, \sin({\bm q}\cdot{\bm
y})\, \chi_{\sigma\sigma}(x-y).
\label{eq:3.2c}
\ee
\ese
The physical interpretation of $S_{\text{eff}}$ is that $S_0$ contains the
smectic order in a mean-field approximation, whereas the second term on the
right-hand side of Eq.\ (\ref{eq:3.2a}) takes into account fluctuations that
can be described as an exchange of smectons between electrons.

We now define $\psi_{\pm}(p) \equiv \psi(p\pm q)$, and $\bar\psi_{\pm}(p)
\equiv \bar\psi(p\pm q)$. If we use a nearly-free electron model for ${\tilde
S}_0$,\cite{NFE_footnote} this allows us to write
\bse
\label{eqs:3.3}
\be
S_{\text{eff}}[\bar\psi,\psi] = S_0[\bar\psi,\psi] +
S_{\text{int}}[\bar\psi,\psi],
\label{eq:3.3a}
\ee
where
\bea
S_0[\bar\psi,\psi] &=& \sum_p \sum_{\sigma=\pm} G_{\sigma}^{-1}(p)\,
\bar\psi_{\sigma}(p)\,\psi_{\sigma}(p)
\nonumber\\
&&\hskip -60pt + \lambda \sum_p \left[\bar\psi_+(p)\psi_-(p+q) +
\bar\psi_-(p)\,\psi_+(p-q)\right],
\label{eq:3.3b}\\
S_{\text{int}}[\bar\psi,\psi] &=& -\frac{\lambda^2}{2}\, \frac{T}{V} \sum_k
\chi_{\sigma\sigma}(k)\,\left[\delta n_{+-}(k-q)\right.
\nonumber\\
&&\hskip -75pt \left. - \delta n_{-+}(k+q)\right]\, \left[\delta n_{+-}(-k-q) -
\delta n_{-+}(-k+q)\right].
\label{eq:3.3c}
\eea
Here
\be
G_{\pm}^{-1}(p) = i\omega - \xi_{{\bm p}\pm{\bm q}},
\label{eq:3.3d}
\ee
with $i\omega$ a fermionic Matsubara frequency. $\xi_{\bm k} = \epsilon_{\bm k}
- \mu$ with $\mu$ the chemical potential and $\epsilon_{\bm k}$ the
single-fermion energy-momentum relation. We also have defined
\be
\delta n_{\sigma_1\sigma_2}(p) = n_{\sigma_1\sigma_2}(p) - \langle
n_{\sigma_1\sigma_2}(p)\rangle,
\label{eq:3.3e}
\ee
where $n_{\sigma_1\sigma_2}(p) = (T/V) \sum_p {\bar\psi}_{\sigma_1}(p)\,
\psi_{\sigma_2}(p-k)$.
\ese
In Eq.\ (\ref{eq:3.3c}) we have dropped contributions where the
$\chi_{\sigma\sigma}$ appears at wave vectors ${\bm k}\pm 2{\bm q}$, as
$\chi_{\sigma\sigma}$ is soft only at $k=0$.

\begin{figure}[t,b,h]
\vskip -0mm
\includegraphics[width=8.0cm]{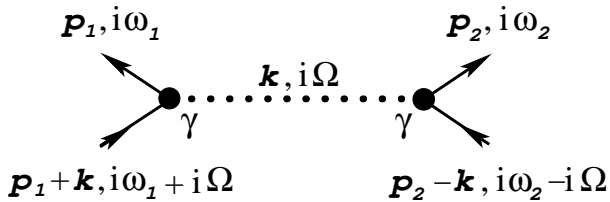}
\caption{The effective quasiparticle interaction, Eq.\ (\ref{eq:3.5c}), due to
smecton exchange. Note that the vertices $\gamma$ depend on the quasiparticle
momenta in addition to the smecton momentum, see Eq.\ (\ref{eq:3.5f}).}
\label{fig:1}
\end{figure}

The action $S_0$ can now be diagonalized by a canonical transformation to
quasiparticle fields $\bar\eta$ and $\eta$ via
\bse
\label{eqs:3.4}
\bea
\psi_-(p) &=& \left[\eta_+(p-q) - \alpha_{{\bm p}-{\bm
q}}\,\eta_-(p-q)\right]/\sqrt{1+\alpha_{{\bm p}-{\bm q}}^2}\ ,
\nonumber\\
\psi_+(p) &=& \left[\eta_-(p) + \alpha_{\bm
p}\,\eta_+(p)\right]/\sqrt{1+\alpha_{\bm p}^2}\ ,
\label{eq:3.4a}
\eea
with
\be
\alpha_{\bm p} = \frac{-1}{2\lambda}\,\left[\xi_{{\bm p}+{\bm q}} - \xi_{\bm p}
+ \sqrt{(\xi_{{\bm p}+{\bm q}} - \xi_{\bm p})^2 + 4\lambda^2}\right],
\label{eq:3.4b}
\ee
\ese
and the same relation between $\bar\psi_{\pm}$ and $\bar\eta_{\pm}$. The
resulting quasiparticle action is very similar to the one for helimagnets
derived in Ref.\ \onlinecite{Kirkpatrick_Belitz_Saha_2008a}. The main
difference is that here the bare quasiparticle Green function $G$, Eq.\
(\ref{eq:3.3d}), depends on the Stoner-band index $\sigma$, where as in the
helimagnon case it does not. The action can be written
\bse
\label{eqs:3.5}
\be
S[\eta,\bar\eta] = S_0[\eta,\bar\eta] + S_{\text{int}}[\eta,\bar\eta],
\label{eq:3.5a}
\ee
where
\bea
S_0 &=& \sum_{p,\sigma} \left[i\omega - \omega_{\sigma}({\bm p})\right]\,
{\bar\eta}_{\sigma}(p)\,\eta(p)
\label{eq:3.5b}\\
S_{\text{int}} &=& -V_0\, \frac{T}{V} \sum_k \chi_{\sigma\sigma}(k)\, \delta
d(k)\, \delta d(-k).
\label{eq:3.5c}
\eea
Here $V_0 = \lambda^2 q^2/8\me^2$, and the resonance frequencies
$\omega_{\pm}({\bm p})$ are given by
\be
\omega_{\pm}({\bm p}) = \frac{1}{2}\,\left[\xi_{{\bm p}+{\bm q}} + \xi_{\bm p}
\pm \sqrt{(\xi_{{\bm p}+{\bm q}} - \xi_{\bm p})^2 + 4\lambda^2}\right]\ .
\label{eq:3.5d}
\ee
$\delta d(k) = d(k) - \langle d(k)\rangle$, with
\bea
d(k) = \sum_p \gamma({\bm k},{\bm p}) \sum_{\sigma} {\bar\eta}_{\sigma}(p)\,
\eta_{\sigma}(p-k),
\label{eq:3.5e}\\
\gamma({\bm k},{\bm p}) = \frac{2\me}{q}\, \frac{\alpha_{\bm p} - \alpha_{{\bm
p}-{\bm k}}}{\sqrt{1+\alpha_{\bm p}^2}\,\sqrt{1+\alpha_{{\bm p}-{\bm k}}^2}}\ .
\label{eq:3.5f}
\eea
\ese
Note that $\gamma({\bm k}\to 0,{\bm p}) \to 0$. This reflects the fact that the
effective interaction described by $S_{\text{int}}$ in Eq.\ (\ref{eq:3.5c})
represents the coupling of electronic density fluctuations to the phase of the
smecton. The phase had no physical significance by itself, and the coupling
must therefore be to the gradient of the phase. The structure of our effective
theory reflects this. The effective interaction is graphically represented in
Fig.\ \ref{fig:1}.

\subsection{Quasiparticle relaxation time}
\label{subsec:III.C}

The effective theory defined by Eqs.\ (\ref{eqs:3.5}) can now be used to
calculate the properties of the quasiparticles by standard means. Electron
correlation functions (which ultimately determine observables such as the
conductivity) can be recovered from quasiparticle ones by means of the
transformation (\ref{eq:3.4a}). We will first focus on the quasiparticle
relaxation time, and then briefly discuss transport properties.

\begin{figure}[b,t]
\vskip -0mm
\includegraphics[width=8.0cm]{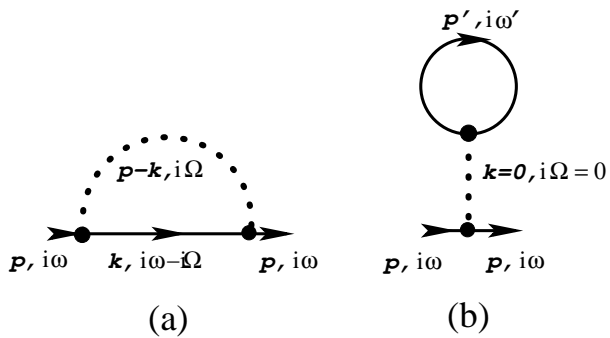}
\caption{The Fock or exchange (a) and Hartree or direct (b) contributions to
 the quasiparticle self energy due to the effective interaction mediated by
 smectons. Only diagram (a) contributes to the relaxation rate, as diagram (b)
 is purely real.}
\label{fig:2}
\end{figure}

The elastic quasiparticle relaxation time $\tau_{\text{el}}$ can be obtained by
adding quenched disorder to the action and calculating the disorder
contribution to the quasiparticle self energy. It does not qualitatively depend
on the dimensionality, and the results given for the $3$-$d$ helimagnon case in
Ref.\ \onlinecite{Kirkpatrick_Belitz_Saha_2008a} apply here as well. The
inelastic quasiparticle relaxation rate $1/\tau$, which is given by the
imaginary part of the quasiparticle self energy due to the interaction
$S_{\text{int}}$, see Fig.\ \ref{fig:2}, in $d=2$ is more interesting. To
linear order in the effective interaction, and keeping only the leading
temperature dependence, the rate averaged over the Fermi surface given by
$\omega_+({\bm k})=0$ can be written
\bea
\frac{1}{\tau} &=& V_0\,\frac{2\gamma_0 \me}{\NF \kF^2}\,\frac{1}{V} \sum_{\bm
p} \frac{\vert p_{\perp}\vert}{\omega_0({\bm p})\,\sinh(\omega_0({\bm p})/T)}\,
\nonumber\\
&&\hskip -20pt \times \frac{1}{V}\sum_{\bm k} \frac{(\partial\alpha_{\bm
k}/\partial k_{\perp})^2}{(1+\alpha_{\bm k})^2}\ \delta\left(\omega_+({\bm
k})\right)\, \delta\left(\frac{\partial\omega_+({\bm k})}{\partial
k_{\perp}}\right)\ .
\nonumber\\
\label{eq:3.6}
\eea
The integrals over ${\bm p}$ and ${\bm k}$ decouple and the temperature
dependence of $1/\tau$ comes from the former. Using Eq.\ (\ref{eq:2.9a}) we see
that it is linear in $T$, with a logarithmically infinite prefactor. The latter
results from the $i\Omega=0$ contribution to the underlying Matsubara frequency
sum; the remainder of the sum leads to a $T\ln T$ behavior. This divergence is
cut off by a variety of effects that have been neglected in the above
treatment. For instance, the underlying lattice structure of a solid (as
opposed to a liquid crystal) breaks the rotational symmetry and produces a term
$\propto {\bm k}_{\perp}^2$ in Eq.\ (\ref{eq:2.9a}), which leads to a $T^2$
behavior of $1/\tau$ at asymptotically low temperatures. Also, screening of the
quasiparticle interaction effectively leads to the same result.
\cite{screening_footnote} There thus is a temperature scale that cuts off the
logarithmic infinity, leaving a $T\ln T$ in the temperature regime where the
current treatment is valid. The remaining question is whether the prefactor
given by the ${\bm k}$-integral in Eq.\ (\ref{eq:3.6}) is nonzero. In general
it is, but this depends on the detailed structure of the Fermi surface. For
instance, for underlying nematic order in $d=2$, which is relevant for stripe
order, it is easy to see that the ${\bm k}$-integral is nonzero if the axes of
the elliptical Fermi surface are not aligned with ${\bm q}$, but vanishes if
they are.\cite{epsilon_k_footnote} We thus conclude that in a $2$-$d$
electronic smectic-C system the quasiparticle relaxation rate displays
non-Fermi-liquid behavior, and its temperature dependence is
\be
1/\tau \propto T\,(\ln T + \text{const.})\quad (\text{clean}, d=2).
\label{eq:3.7}
\ee
The prefactor depends on the value of $q$, and in particular is strongly
dependent on whether $q \kF/\me$ is small or large compared to $\lambda$. This
will be discussed in a future publication. In $d=3$ the corresponding
temperature dependence is $T^{3/2}$, as it is for helimagnons.
\cite{Belitz_Kirkpatrick_Rosch_2006b} For real systems, these results hold in a
pre-asymptotic temperature regime whose size depends on detailed parameter
values, and it will cross over to a different behavior in the true asymptotic
low-temperature regime. Comparing with Ref.\ \onlinecite{Sun_et_al_2008}, who
found a stronger behavior $1/\tau \propto \ln T$, we see that the discrepancy
stems from the coupling of the smectons to the quasiparticles. In Ref.\
\onlinecite{Sun_et_al_2008} the quasiparticles couple to the smecton phase,
rather than to its gradient as they should on physical grounds.

The presence of quenched disorder modifies the above considerations. In the
weak-disorder regime,\cite{Zala_Narozhny_Aleiner_2001,
Kirkpatrick_Belitz_Saha_2008a} $T\gg
\lambda/(\epsilon_{\text{F}}\tau_{\text{el}})^2$, the calculation proceeds in
analogy to Ref.\ \onlinecite{Kirkpatrick_Belitz_Saha_2008a}. The result is
\be
1/\tau \propto T^{1/2}\quad(\text{weak disorder}, d=2)
\label{eq:3.8}
\ee
in $d=2$, and $1/\tau \propto T$ in $d=3$.

\subsection{Transport coefficients}
\label{subsec:III.D}

The above results pertain to the quasiparticle relaxation time, which is not
easy to observe directly. Various transport coefficients depend on relaxation
times that are generally different from the quasiparticle one, and that are
much harder to calculate. For instance, the Boltzmann equation for the
electrical conductivity $\sigma$, with the scattering treated in Born
approximation, leads to a transport relaxation rate that is, for the current
problem, weaker by one power of temperature than the quasiparticle
rate.\cite{Belitz_Kirkpatrick_Rosch_2006b} Technically, the electrical
transport relaxation rate is given by Eq.\ (\ref{eq:3.6}) with an additional
factor proportional to ${\bm p}_{\perp}^{\ 2}$ in the integrand of the ${\bm
p\,}$-integral. This leads to $\sigma \propto 1/T^2$ in $d=2$, and $\sigma
\propto 1/T^{5/2}$ in $d=3$.

The situation is different, however, for the thermal conductivity $\kappa$: The
temperature dependence of $\kappa/T$ is given by the quasiparticle relaxation
rate.\cite{Wilson_1954, Ziman_1972} The physical reason is that an electric
current can relax only by the electrons changing direction, since the
electron's electric charge is conserved. In the calculation of the relaxation
time, this leads to a geometric factor that weighs backscattering more strongly
than forward scattering, and this is manifested in the additional factor of
${\bm p}_{\perp}^{\ 2}$ in the integrand. An electron's energy is not
conserved, however, in an inelastic scattering process, and hence this
geometric factor is absent in the calculation of the leading temperature
dependence of the thermal transport coefficient.\cite{Ziman_1972,
Ashcroft_Mermin_1976} In the current problem, this leads to
\be
\kappa/T \propto 1/T\ln T\quad (\text{clean}, d=2)
\label{eq:3.9}
\ee
for clean systems in $d=2$. In $d=3$, the corresponding temperature dependence
is $T^{-3/2}$. The Wiedemann-Franz law is thus violated, as it is in the case
of electron-phonon scattering, and the Lorenz ratio, defined by $L =
\kappa/T\sigma$, is proportional to $T\ln T$. In weakly disordered systems, the
temperature dependence of $\kappa/T$ is governed by Eq.\ (\ref{eq:3.8}), but
one needs to take into account the residual value $\kappa_{\text{r}}$ of the
thermal conductivity. Since the residual values of the transport coefficients
are determined by elastic scattering processes, the latter is related to the
residual electrical conductivity $\sigma_{\text{r}}$ by the Wiedemann-Franz law
$\kappa_{\text{r}}/T\sigma_{\text{r}} = L_{\text{r}}$ with the Lorenz ratio
$L_{\text{r}} = \pi^2 \kB^2/3 e^2$ a constant.

\section{Summary, and Conclusion}
\label{sec:IV}

In summary, we have determined the Goldstone modes and their properties in
electronic smectics or stripe phases. In an isotropic model system the soft
modes (``smectons'') are precisely analogous to those in both smectic and
cholesteric classical liquid crystals. Their wave-vector dependence differs
(albeit not in a scaling sense) from that of the helimagnons in helical
magnets, which are analogous to classical cholesteric liquid crystals. This
difference is due to the fact that spin dynamics are different from density
dynamics.

In $d=2$, the smectons contribute a term proportional to $T^{3/2}$ to the
specific heat, and the quasiparticle relaxation rate $1/\tau$ as well as
$T/\kappa$, with $\kappa$ the heat conductivity, are proportional to $T\ln T$
in clean systems. In weakly disordered systems, the corresponding leading
temperature dependence is given by $T^{1/2}$. In $d=3$, the corresponding
temperature dependencies are $T^2$ (for the specific heat), $T^{3/2}$ (for the
relaxation rate and $T/\kappa$ in clean systems), and $T$ (for the relaxation
rate and $T/\kappa$ in the weak-disorder regime), respectively. In the
weak-disorder regime, the leading temperature dependencies of $1/\tau$,
$1/\sigma$, and $T/\kappa$. The leading temperature dependence of the
electrical resistivity $1/\sigma$ is weaker than that of $1/\tau$ by one power
of $T$ in clean systems, and the same as that of $1/\tau$ in weakly disordered
ones, respectively. Qualititively, all of these results also hold for the
exchange of helimagnons between electrons in helical
magnets,\cite{Belitz_Kirkpatrick_Rosch_2006b, Kirkpatrick_Belitz_Saha_2008b}
and they are summarized in Table\ \ref{table:1}.
\begin{table}[t,b,h]
\begin{ruledtabular}
\begin{tabular}{ccccccccc}
          & \vline\,\vline &       & clean  &       & \vline\,\vline &       & weak disorder &
          \\
\hline
          & \vline\,\vline & $d=2$ & \vline & $d=3$ & \vline\,\vline & $d=2$ & \vline & $d=3$\\
\hline \hline
          & \vline\,\vline &           & \vline &           & \vline\,\vline &           & \vline & \\
C         & \vline\,\vline & $T^{3/2}$ & \vline & $T^2$     & \vline\,\vline & $T^{3/2}$ & \vline & $T^2$ \\
$1/\tau$  & \vline\,\vline & $T\ln T$  & \vline & $T^{3/2}$ & \vline\,\vline & $T^{1/2}$ & \vline & $T$ \\
$T/\kappa$& \vline\,\vline & $T\ln T$  & \vline & $T^{3/2}$ & \vline\,\vline & $T^{1/2}$ & \vline & $T$ \\
$1/\sigma$& \vline\,\vline & $T^2$     & \vline & $T^{5/2}$ & \vline\,\vline & $T^{1/2}$ & \vline & $T$ \\
\end{tabular}
\end{ruledtabular}
\caption{Leading smecton contributions to the temperature dependencies of the
 specific heat ($C$), the quasiparticle relaxation rate ($1/\tau$), the heat
 conductivity ($\kappa$), and the electrical conductivity ($\sigma$). The same
 results hold for the helimagnon contributions in helical magnets. See the text
 for additional information.}
\label{table:1}
\end{table}

We conclude with some speculations pertaining to the electrical conductivity.
While the standard weak-coupling treatment of the Boltzmann equation yields a
resistivity $\rho \propto T^2$ in $2$-$d$ clean systems as mentioned above, it
is conceivable that in a strongly correlated electron system mode-mode coupling
effects mix the various time scales and lead to a single relaxation time. It is
currently not known whether this hypothesis is correct, or what it takes at a
technical level to demonstrate it, but it provides a possible mechanism for
producing an electrical resistivity that is linear in $T$ in $2$-$d$ or
quasi-$2$-$d$ systems.

\acknowledgments

This work was supported by the NSF under grant Nos. DMR-05-29966 and
DMR-05-30314.

\vskip -0mm

\begin{thebibliography}{39}
\bibitem[{sof()}]{soft_mode_footnote}
\bibinfo{note}{One might ask how a massive mode (the plasmon) can modify a soft
  mode by coupling to it. The answer lies in the fact that the modification
  (namely, the factor of ${\bm k}_{\perp}^2/{\bm k}^2$ in Eq.\
  (\ref{eq:2.10b})), although it introduces a strong angular dependence of the
  resonance frequency, is of $O(1)$ in a scaling sense, and hence cannot be
  ruled out by power counting.}
\bibitem[{not()}]{notation_footnote}
\bibinfo{note}{In our notation, $a\propto b$ and $as\sim b$ mean ``$a$ is
  proportional to $b$'' and ``$a$ scales as $b$'', respectively.}
\bibitem[{NFE()}]{NFE_footnote}
\bibinfo{note}{We note that all interactions that are necessary to create and
  stabilize smectic order are already implicit in the second contribution to
  the action $S_0$, Eq.\ (\ref{eq:3.2b}), i.e., in the existence of a nonzero
  Stoner gap $\lambda$. The residual interactions included in ${\tilde S}_0$
  are not expected to have a qualitative influence on our results, and we
  neglect them.}
\end{thebibliography}

\end{document}